\documentclass[a4paper,onecolumn,fleqn,usenatbib]{mnras}
\usepackage{newtxtext,newtxmath}
\usepackage[T1]{fontenc}
\usepackage{ae,aecompl}
\usepackage{graphicx}

\newcommand       \be           {\begin{equation}}
\newcommand       \ee           {\end{equation}}

\title[Barnett relaxation in non-symmetric grains]
{Barnett relaxation in non-symmetric grains}
\author[E. Kolasi \&  J. C. Weingartner]
{Erald Kolasi$^{1}$\thanks{E-mail: ekolasi@masonlive.gmu.edu; jweinga1@gmu.edu}
and  Joseph C. Weingartner$^{1}$\\
$^{1}$Department of Physics and Astronomy, George Mason University, 
4400 University Drive, Fairfax, VA 22030, USA}

\date{Accepted XXX. Received YYY; in original form ZZZ}

\pubyear{2015}

\begin{document}
\label{firstpage}
\pagerange{\pageref{firstpage}--\pageref{lastpage}}
\maketitle

\begin{abstract}

Barnett relaxation, first described by Purcell in 1979, appears to play a
major role in the alignment of grains with the interstellar magnetic field.
In 1999, Lazarian and Draine proposed that Barnett relaxation and its relative,
nuclear relaxation, can induce grains to flip. If this thermal flipping is
rapid, then the dynamical effect of torques that are fixed relative to the
grain body can be greatly reduced. To date, detailed studies of Barnett
relaxation have been confined to grains exhibiting dynamic symmetry.
In 2003, Weingartner argued that internal relaxation cannot induce flips in
any grains, whether they exhibit dynamic symmetry or not. In this work, we
develop approximate expressions for the dissipation rate and diffusion
coefficient for Barnett relaxation. We revisit the issue of internally 
induced thermal flipping, finding that it cannot occur for grains with
dynamic symmetry but does occur for grains lacking dynamic symmetry. 

\end{abstract}


\section{Introduction}

The discovery of starlight polarization nearly 70 years ago
\citep{Hall49,Hall-Mikesell49,Hiltner49a,Hiltner49b} revealed that
interstellar dust grains are nonspherical and aligned with the interstellar
magnetic field. Yet we still do not have a complete, unambiguous theory of
the alignment mechanism.

In seminal work, \citet{Purcell79} identified the
process of ``Barnett dissipation'', in which rotational kinetic energy
is converted into vibrational energy in paramagnetic grains. If a grain
rotates steadily around a principal axis, then there is no Barnett dissipation.
Otherwise, the grain's angular velocity $\bmath{\omega}$ is not constant
as observed in the grain-body frame (i.e.~the non-inertial frame in which the 
grain does not rotate). The microscopic spins in the paramagnetic material
tend to align along the ``Barnett-equivalent magnetic field''
$\mathbfit{B}_{\mathrm{BE}} = \bmath{\omega}/\gamma_g$ ($\gamma_g$ is the
gyromagnetic ratio of the microscopic spins), but with a lag. As a result,
energy is dissipated into heat, as in magnetic resonance.

\citet{LR97} examined the inverse process, in which thermal
fluctuations convert vibrational energy into rotational kinetic energy,
deriving an expression for the diffusion coefficient that appears in the
Fokker-Planck and Langevin equations describing Barnett relaxation
(i.e.~the combination of both dissipation and fluctuations).

Barnett relaxation is an internal process, unrelated to any external torques
or other influences. Thus, the grain angular momentum $\mathbfit{J}$ remains
constant under Barnett relaxation. \citet{LD99a} introduced
the concept of ``thermal flipping'', in which the sign of
$\mathbfit{J} \bmath{\cdot} \bmath{\hat{a}}_1$
changes as a result of a Barnett fluctuation; $\bmath{\hat{a}}_1$ is the
principal axis of greatest moment of inertia. 
Thermal flipping can potentially have important implications for grain
alignment, due to another major insight in \citet{Purcell79}, namely the
existence of systematic torques fixed relative to the grain body. These
torques can spin the grain up to rapid rotation, making them impervious to
disalignment via random collisions with gas particles. Each grain flip
reverses the direction of the systematic torque in space (i.e.~relative to
an inertial frame). If flips occur in rapid succession, then the average
systematic
torque is zero and the grain may not be spun up after all. \citet{LD99a}
refer to this condition as ``thermal trapping''.  Noting the
potential importance of nuclear paramagnetism in addition to electron
paramagnetism, \citet{LD99b} examined ``nuclear relaxation'' and
concluded that grains that are responsible for the observed starlight
polarization (i.e.~with sizes $< 1 \, \mu$m) are likely thermally trapped.

\citet{W09} revisited the treatment of Barnett relaxation in \citet{LR97}
and found a revised expression for the diffusion coefficient.
With this revision, thermal flipping was not possible. Shortly thereafter,
\citet{HL09} found that flips resulting from collisions with gas
particles likely occur sufficiently rapidly for grains to become trapped.

For simplicity, treatments of Barnett relaxation have focused on oblate grains
with dynamic symmetry. That is, the moments of inertia associated with the
principal axes $\bmath{\hat{a}}_1$, $\bmath{\hat{a}}_2$, and
$\bmath{\hat{a}}_3$ have $I_2 = I_3$ and $I_1 > I_2$.
It seems unlikely that real interstellar grains would be characterized by
dynamic symmetry. Thus, in this work, we examine Barnett relaxation for
non-symmetric grains, with three different principal moments of inertia.

Although \citet{W09} computed the diffusion coefficient only for
symmetric grains, he described general considerations that implied that
thermal flipping would be impossible also for non-symmetric grains.
However, we find that thermal flipping is possible for non-symmetric grains.
In section \ref{sec:symmetric-grains}, we briefly summarize and revisit the
analysis in \citet{W09}. We conclude that thermal flipping is not
possible for symmetric grains, but not necessarily for the reason \citet{W09}
identified. The argument applies only to symmetric grains and is mute
regarding non-symmetric grains. 
In sections \ref{sec:barnett-diss} and \ref{sec:diffusion-coeff}, respectively,
we develop approximate expressions for the Barnett dissipation rate and
diffusion coefficient for non-symmetric grains. Section
\ref{sec:simulations} describes the results of simulations of Barnett
relaxation for a few shapes, exhibiting rapid thermal flipping. 
Section \ref{sec:summary} summarizes the results. 

\section{Symmetric Grains} \label{sec:symmetric-grains}

For an oblate grain with dynamic symmetry, the solution of Euler's equation is
simple. The component of the angular velocity along $\bmath{\hat{a}}_1$, the
principal axis of greatest moment of inertia, is constant with magnitude
\be
\omega_{\parallel} = \frac{J}{I_1} \, \cos \gamma ,
\ee
where $J$ is the angular momentum and $\gamma$ is the angle between
$\mathbfit{J}$ and $\bmath{\hat{a}}_1$. There is also a component in the
$\bmath{\hat{a}}_2$--$\bmath{\hat{a}}_3$ plane with magnitude
\be
\omega_{\perp} = \frac{J}{I_2} \, \sin \gamma
\ee
that rotates with angular speed
\be
\omega_{\mathrm{rot}} = \frac{J (I_1 - I_2)}{I_1 I_2} \, \cos \gamma .
\ee

Since the rotating component of the Barnett-equivalent field
$\mathbfit{B}_{\mathrm{BE, rot}} = \bmath{\omega}_{\perp}/\gamma_g$, a direct analogy
with the standard treatment of magnetic resonance offers an immediate expression
for the rate at which the grain's rotational kinetic energy is dissipated:
\be
\left( \frac{dE}{dt} \right)_{\mathrm{Bar}} = - V \chi^{\prime \prime}
B_{\mathrm{BE, rot}}^2 \omega_{\mathrm{rot}} ,
\label{eq:diss-symm}
\ee
where $V$ is the grain volume and $\chi^{\prime \prime}$ is the imaginary
component of the magnetic susceptibility. When introducing this result,
\citet{Purcell79} adopted the low-frequency susceptibility
\be
\chi^{\prime \prime} \approx \chi_0 \omega_{\mathrm{rot}} T_2 ,
\ee
where $\chi_0$ is the static susceptibility and $T_2$ is the spin-spin
relaxation time. This expression, as well as a more general result for
arbitrary frequencies, can be inferred from the Bloch equations. 

It is convenient to introduce a dimensionless measure of the rotational
kinetic energy,
\be
q = \frac{2 I_1 E}{J^2} .
\ee
From equation (\ref{eq:diss-symm}), the corresponding dissipation rate is
\be
A(q) = \frac{dq}{dt} = - \tau_{\mathrm{Bar}}^{-1} (q-1) (r_2 -q)
\label{eq:drift-symm}
\ee
with
\be
\tau_{\mathrm{Bar}} = \frac{\gamma_g^2 I_1 I_2^2}{2 \chi_0 V T_2 J^2} 
\ee
and $r_2 = I_1/I_2$. This expression for $A(q)$ is the drift coefficient in
the Fokker-Planck and Langevin equations.

To find the diffusion coefficient $D(q)$, \citet{LR97} and
\citet{W09} demanded that the probability current
\be
S(q) = A(q) f(q) - \frac{1}{2} \frac{\mathrm{d}[f(q) D(q)]}{\mathrm{d}q}
\label{eq:prob-current}
\ee
vanish for all $q$ when thermal equilibrium applies. Here $f(q)$ is the
probability density. That is, $f(q) \mathrm{d}q$ is the probability that the
dimensionless energy lies between $q$ and $q+\mathrm{d}q$. Thus,
\be
D(q) = \frac{1}{f_{\rm TE}(q)} \left[ D(1) f_{\rm TE}(1) + 2 
\int_1^q A(q^{\prime}) f_{\rm TE}(q^{\prime}) \mathrm{d}q^{\prime} \right] ,
\label{eq:diff-coeff-as-integral}
\ee
where $f_{\rm TE}(q)$ denotes the thermal-equilibrium distribution.

For an oblate grain with dynamic symmetry,
\be
q = 1 + (r_2 -1) \sin^2 \gamma
\label{eq:q-vs-gamma}
\ee
and $1 \le q \le r_2$. As seen in equation (\ref{eq:q-vs-gamma}), 
the rotational kinetic energy (parametrized by $q$ for constant grain
angular momentum $\mathbfit{J}$) is unchanged if
$\gamma \rightarrow \pi - \gamma$.
In other words, $q$ does not depend on the sign of $\cos \gamma$. So to fully
characterize the grain's rotational state, one must specify $\mathbfit{J}$,
$q$, and the sign of $\cos \gamma$ (which is equal to the sign of
$\mathbfit{J} \bmath{\cdot} \bmath{\hat{a}}_1$). \citet{WD03} denoted
the latter as the ``flip state'' of the grain. 
In order for a grain to flip, $q$ must increase to $r_2$, where
$\gamma = \pi/2$ and $\cos \gamma = 0$, and then return to $q < r_2$, but with
a change in sign of $\cos \gamma$ (i.e.~in the opposite flip state). 

\citet{W09} attempted to find the constant of integration in
equation (\ref{eq:diff-coeff-as-integral}) by demanding that fluctuations
cease to contribute to $S(q)$ in the limit that the dust temperature
$T_d \rightarrow 0$. This must be true regardless of the form of $f(q)$.
In order to be normalizable, $f(q)$ may diverge at any value $q_0$, but the
divergence must be shallower than $f(q) \propto |q_0 -q|^{-1}$. 
Defining
\be
b = \frac{J^2}{2 I_1 k_B T_d} 
\ee
(with $k_B$ Boltzmann's constant), 
it must be the case that, for any $q$,
\be
\lim_{b^{-1} \rightarrow 0} \frac{\mathrm{d}[f(b, q) D(b,q)]}{\mathrm{d}q} = 0 .
\ee
\citet{W09} claimed that the stronger condition  
\be
\label{eq:lim}
\lim_{(b^{-1}, r_2-q) \rightarrow (0,0)} \frac{\mathrm{d}[f(b, q) D(b,q)]}
{\mathrm{d}q} = 0 
\ee
must be satisfied at $q=r_2$. This will only hold for all possible forms
of $f(b,q)$ if $D(b,q)$ falls off at least as fast as $(r_2-q)^2$ near
$q=r_2$, implying that both $D(q)$ and $\mathrm{d}D/\mathrm{d}q$ vanish at
$q=r_2$. These results, along with the fact that $A(q)$ vanishes at $q=r_2$,
make $q=r_2$ a natural boundary \citep{Gardiner04}. That is,
assuming the condition in equation (\ref{eq:lim}) is enforced, 
the grain cannot evolve to $q=r_2$ if it starts at $q \neq r_2$. 

For non-symmetric grains with $I_1 > I_2 > I_3$, $1 \le q \le r_3$, where
$r_3 = I_1/I_3$. Of course, $r_3 > r_2$. As described in section
\ref{sec:simulations}, for the grain to flip, it must start
with $q < r_2$, evolve to $q > r_2$, and then evolve to $q < r_2$ again.
If $q=r_2$ is a natural
boundary, then, since the argument in the above paragraph did not make use
of symmetry in any way, even non-symmetric grains cannot undergo thermal
flipping.

However, we cannot see any reason why the stronger condition in
equation (\ref{eq:lim}) must be satisfied. Furthermore, nothing in the
argument is restricted to $q=r_2$. It could be applied as well to any other
value of $q$, making them all natural boundaries, which is absurd.

Even if $q=r_2$ is not a natural boundary, 
a symmetric grain never reaches $q=r_2$ for the same reason that it never
reaches $q=1$, which is certainly not a natural boundary. The probability that
$q$ evolves to exactly $1$ or exactly $r_2$ is infinitesimal. Since $q$ must
reach exactly $r_2$ for the grain to undergo a thermal flip, thermal flipping
is not possible for symmetric grains. In non-symmetric grains, $q$ must
simply evolve past $q=r_2$ in order to undergo a thermal flip (assuming
$q=r_2$ is not a natural boundary, to be addressed in section
\ref{sec:diffusion-coeff}).


The conclusion that thermal flipping is impossible for symmetric grains
relies on the fact that the rotational kinetic energy $E$ is independent of
the flip state and the assumption that the fluctuations are in $E$
(or equivalently $q$ for constant $\mathbfit{J}$). \citet{LR97}
did not incorporate this assumption in their model of Barnett relaxation,
enabling \citet{LD99a} to conclude that thermal flipping occurs
for symmetric grains. When the evolution of the
grain's rotational kinetic energy is affected by collisions with gas
atoms, which also change the grain's angular momentum, this assumption no
longer applies and flipping is possible. A detailed microphysical model of
Barnett relaxation is needed to confirm or deny the above assumption and
provide an unambiguous description of thermal flipping. 

For oblate grains with dynamic
symmetry, the equilibrium probability density is
\be
f_{\mathrm{TE}}(q) \propto \exp(-bq) \, (r_2 - q)^{-1/2}
\label{eq:f-TE-symm}
\ee
\citep[see, e.g.,][]{W09}.
Substituting for $A(q)$ and $f_{\mathrm{TE}}(q)$ from equations
(\ref{eq:drift-symm}) and (\ref{eq:f-TE-symm}) in equation
(\ref{eq:diff-coeff-as-integral}) and integrating,
\be
\label{eq:D}
b^2 \tau_{\mathrm{Bar}} D(q, b) = [3 + 2 b (q-1)] (r_2 - q) + C(b)
(r_2 - q)^{1/2} \exp(bq)
- b^{-1/2} [3 + 2 b (r_2 -1)] (r_2 - q)^{1/2} F_D \left(\sqrt{b (r_2 -q)}
\right)
\ee
where
\be
F_D(x) = \exp(-x^2) \int_0^x \exp(y^2) \, \mathrm{d}y
\ee
is Dawson's integral and
\be
C(b) = b^2 \exp(-b) (r_2 -1)^{-1/2} \tau_{\mathrm{Bar}} D(q=1, b) - 3 \exp(-b)
(r_2 -1)^{1/2}
+ b^{-1/2} [3 + 2 b (r_2 -1)] \exp(-b) F_D \left(\sqrt{b (r_2 -1)}
\right) .
\ee
When $r_2 - q \ll 1$,
\be
b^2 \tau_{\mathrm{Bar}} D(q, b) \approx C(b) \exp(bq) (r_2 - q)^{1/2} +
\frac{4}{3} b^2 (r_2 -1) (r_2 -q)^2 .
\ee
Applying the condition in equation (\ref{eq:lim}), \citet{W09}
concluded that $C(b) = 0$.

Dispensing with this condition, we find that
$C(b)$ cannot be uniquely determined by examining the low-temperature
limit (i.e.~$b \rightarrow \infty$). Rather, if $C(b)$ is not identically
zero, then it must fall off at least as fast as $b^2 \exp(-br_2)$ as
$b \rightarrow \infty$. We have not been able to identify a general argument
that can uniquely determine $C(b)$; perhaps this is only possible with a
detailed microphysical model.

\section{Barnett Dissipation in Non-Symmetric Grains} \label{sec:barnett-diss}

\subsection{Euler's equations}

Consider a non-symmetric grain with principal moments of inertia
$I_1 > I_2 > I_3$ and denote $r_2 = I_1/I_2$ and $r_3 = I_1/I_3$. When
$1 < q < r_2$, the solution
of Euler's equations for the angular velocity $\bmath{\omega}$ is given in
terms of Jacobi elliptic functions as 
\be
\label{eq:omega1-low-q}
\omega_1 = \pm \frac{J}{I_1} \left( \frac{r_3-q}{r_3-1} \right)^{1/2} 
\mathrm{dn}(\omega_{\mathrm{rot}} t, k^2) ,
\ee
\be
\omega_2 = - \frac{J}{I_1} r_2 \left( \frac{q-1}{r_2-1} \right)^{1/2} 
\mathrm{sn}(\omega_{\mathrm{rot}} t, k^2) ,
\ee
\be
\omega_3 = \pm \frac{J}{I_1} r_3 \left( \frac{q-1}{r_3-1} \right)^{1/2} 
\mathrm{cn}(\omega_{\mathrm{rot}} t, k^2) ,
\label{eq:omega3-low-q}
\ee
where
\be
\label{eq:k2}
k^2 = \frac{(r_3-r_2) (q-1)}{(r_2-1) (r_3-q)}
\ee
and
\be
\omega_{\mathrm{rot}} = \frac{J}{I_1} \left[ (r_2-1) (r_3-q) \right]^{1/2} .
\ee
The grain is in the positive flip state with respect to $\bmath{\hat{a}}_1$
(i.e.~$\mathbfit{J} \bmath{\cdot} \bmath{\hat{a}}_1 > 0$) when the plus sign
is chosen
in both equations (\ref{eq:omega1-low-q}) and (\ref{eq:omega3-low-q}). It is
in the negative flip state with respect to $\bmath{\hat{a}}_1$ when the minus
sign is chosen in both of those cases. We adopt the same conventions for the
Jacobi elliptic functions as \citet{WD03}. 

When $r_2 < q < r_3$,
\be
\label{eq:omega1-high-q}
\omega_1 = \pm \frac{J}{I_1} \left( \frac{r_3-q}{r_3-1} \right)^{1/2} 
\mathrm{cn}(\omega_{\mathrm{rot}} t, k^{-2}) ,
\ee
\be
\omega_2 = - \frac{J}{I_1} r_2 \left( \frac{r_3 - q}{r_3-r_2} \right)^{1/2} 
\mathrm{sn}(\omega_{\mathrm{rot}} t, k^{-2}) ,
\ee
\be
\omega_3 = \pm \frac{J}{I_1} r_3 \left( \frac{q-1}{r_3-1} \right)^{1/2} 
\mathrm{dn}(\omega_{\mathrm{rot}} t, k^{-2}) ,
\label{eq:omega3-high-q}
\ee
with
\be
\omega_{\mathrm{rot}} = \frac{J}{I_1} \left[ (r_3-r_2) (q-1) \right]^{1/2} .
\ee
The grain is in the positive flip state with respect to $\bmath{\hat{a}}_3$
(i.e.~$\mathbfit{J} \bmath{\cdot} \bmath{\hat{a}}_3 > 0$) when the plus sign
is chosen
in both equations (\ref{eq:omega1-high-q}) and (\ref{eq:omega3-high-q}). It is
in the negative flip state with respect to $\bmath{\hat{a}}_3$ when the minus
sign is chosen in both of those cases.

As for symmetric grains, the Barnett equivalent field
$\mathbfit{B}_{\mathrm{BE}} = \bmath{\omega}/\gamma_g$. 

\subsection{Modified Bloch equations}

Since the Bloch equations do not accomodate dissipation in the limit of
zero biasing field, we adopt the modified Bloch equations of
\citet{Wangsness56}. These are
\be
\frac{\mathrm{d}M_x}{\mathrm{d}t} = \gamma_g \left( M_y B_z - M_z B_y \right) - 
\frac{M_x - \chi_0 B_x}{T_2} ,
\ee
\be
\frac{\mathrm{d}M_y}{\mathrm{d}t} = \gamma_g \left( M_z B_x - M_x B_z \right) - 
\frac{M_y - \chi_0 B_y}{T_2} ,
\ee
\be
\frac{\mathrm{d}M_z}{\mathrm{d}t} = \gamma_g \left( M_x B_y - M_y B_x \right) - 
\frac{M_z - \chi_0 B_z}{T_1} , 
\ee
where $\mathbfit{M}$ is the magnetization, 
$T_1$ is the spin-lattice relaxation time, and the other quantities were
introduced in section \ref{sec:symmetric-grains}. In contrast
to typical magnetic resonance experiments, where $B_z$ is constant, all
three Barnett equivalent fields are time-variable for the case of a 
nonsymmetric grain.  It is not clear how to further modify the Bloch 
equations for this case, so we will simply set $T_1 = T_2$.  

Taking $(\bmath{\hat{x}},\bmath{\hat{y}},\bmath{\hat{z}})$ in the modified
Bloch equations to lie along
$(\bmath{\hat{a}}_2,\bmath{\hat{a}}_3,\bmath{\hat{a}}_1)$ and introducing
dimensionless variables
\be
t^{\prime} = t/T_2 ,
\ee 
\be
T_2^{\prime} = \omega_{\mathrm{rot}} T_2 ,
\ee
\be
\mathbfit{B}^{\prime} = \frac{\gamma_g \mathbfit{B}}{\omega_{\mathrm{rot}}^2 T_2} ,
\ee
and
\be
\mathbfit{M}^{\prime} = \frac{\gamma_g \mathbfit{M}}{\chi_0 \omega_{\mathrm{rot}}} ,
\ee
the modified Bloch equations become
\be
\frac{\mathrm{d}M^{\prime}_2}{\mathrm{d}t^{\prime}} = (T^{\prime}_2)^2 \left(
M^{\prime}_3 B^{\prime}_1 
- M^{\prime}_1 B^{\prime}_3 \right) - M^{\prime}_2 + T^{\prime}_2 B^{\prime}_2 ,
\label{eq:modified-bloch-2}
\ee
\be
\frac{\mathrm{d}M^{\prime}_3}{\mathrm{d}t^{\prime}} = (T^{\prime}_2)^2 \left(
M^{\prime}_1 B^{\prime}_2 
- M^{\prime}_2 B^{\prime}_1 \right) - M^{\prime}_3 + T^{\prime}_2 B^{\prime}_3 ,
\ee
\be
\frac{\mathrm{d}M^{\prime}_1}{\mathrm{d}t^{\prime}} = (T^{\prime}_2)^2 \left(
M^{\prime}_2 B^{\prime}_3 
- M^{\prime}_3 B^{\prime}_2 \right) - M^{\prime}_1 + T^{\prime}_2 B^{\prime}_1 .
\label{eq:modified-bloch-1}
\ee
We will consider the low-frequency limit, with $T_2^{\prime} \ll 1$.

\subsection{The case that $1 < q < r_2$}

From equations (\ref{eq:omega1-low-q})--(\ref{eq:omega3-low-q}), 
when $1 < q < r_2$, 
\be
B_1^{\prime} = (T^{\prime}_2)^{-1} c_1 \, \mathrm{dn}(T^{\prime}_2 t^{\prime}, k^2) ,
\ee
\be
B_2^{\prime} = (T^{\prime}_2)^{-1} c_2 \, \mathrm{sn}(T^{\prime}_2 t^{\prime}, k^2) ,
\ee
\be
B_3^{\prime} = (T^{\prime}_2)^{-1} c_3 \, \mathrm{cn}(T^{\prime}_2 t^{\prime}, k^2) ,
\ee
with 
\be
c_1 = \left[(r_2-1) (r_3-1) \right]^{-1/2} ,
\ee
\be
c_2 = - \frac{r_2}{r_2-1} \left( \frac{q-1}{r_3-q} \right)^{1/2} ,
\ee
\be
c_3 = r_3 \left[ \frac{q-1}{(r_2-1) (r_3-1) (r_3-q)} \right]^{1/2} .
\ee

The solution to the modified Bloch equations is
\be
M^{\prime}_1 = c_1 \, \mathrm{dn}(T^{\prime}_2 t^{\prime}, k^2) + 
T_2^{\prime} k^2 c_1 \, \mathrm{sn}(T^{\prime}_2 t^{\prime}, k^2) \,
\mathrm{cn}(T^{\prime}_2 t^{\prime}, k^2) 
+ \left( T_2^{\prime} \right)^2 \mathrm{dn}(T^{\prime}_2 t^{\prime}, k^2)
\left\{ k^2 c_1 \left[ 2 \, \mathrm{sn}^2(T^{\prime}_2 t^{\prime}, k^2) -1 \right] -
c_2 c_3 \right\} + O[(T_2^{\prime})^3] ,
\ee
\begin{multline}
M^{\prime}_2 = c_2 \, \mathrm{sn}(T^{\prime}_2 t^{\prime}, k^2) - 
T_2^{\prime} c_2 \, \mathrm{cn}(T^{\prime}_2 t^{\prime}, k^2) \,
\mathrm{dn}(T^{\prime}_2 t^{\prime}, k^2) \\
+ \left( T_2^{\prime} \right)^2 \mathrm{sn}(T^{\prime}_2 t^{\prime}, k^2)
\left\{ \left( 1 - k^2 \right) c_1 c_3 - c_2 \left[ 2 \,
\mathrm{dn}^2(T^{\prime}_2 t^{\prime}, k^2) - \left( 1 - k^2 \right) \right]
\right\} + O[(T_2^{\prime})^3] ,
\end{multline}
\be
M^{\prime}_3 = c_3 \, \mathrm{cn}(T^{\prime}_2 t^{\prime}, k^2) + 
T_2^{\prime} c_3 \, \mathrm{sn}(T^{\prime}_2 t^{\prime}, k^2) \,
\mathrm{dn}(T^{\prime}_2 t^{\prime}, k^2)
+ \left( T_2^{\prime} \right)^2 \mathrm{cn}(T^{\prime}_2 t^{\prime}, k^2)
\left\{ c_1 c_2 + c_3 \left[ 1 - 2 \, \mathrm{dn}^2(T^{\prime}_2
t^{\prime}, k^2) \right] \right\} + O[(T_2^{\prime})^3] ,
\ee
where $O[(T_2^{\prime})^3]$ denotes terms of order $(T_2^{\prime})^3$ and higher.

The instantaneous absorbed power per unit volume is
$\mathbfit{B} \bmath{\cdot} \mathrm{d}\mathbfit{M}/\mathrm{d}t$.
The period of the motion is $4 \omega_{\mathrm{rot}}^{-1} K(k^2)$, where
$K(k^2)$ is the complete elliptic integral of the first kind, 
\be
K(k^2) = \int_0^{\pi/2} \mathrm{d}\theta (1-k^2 \sin^2 \theta)^{-1/2} .
\ee
Thus, the grain's rotational kinetic energy is dissipated at rate
\be
\frac{\mathrm{d}E}{\mathrm{d}t} = - \frac{V \omega_{\mathrm{rot}}}{4 K(k^2)} 
\int_0^{4 \omega_{\mathrm{rot}}^{-1} K(k^2)} \mathrm{d}t \, \mathbfit{B} \bmath{\cdot}
\frac{\mathrm{d}\mathbfit{M}}{\mathrm{d}t}
= - \frac{\chi_0 V T_2 \omega_{\mathrm{rot}}^4}{4 K(k^2) \gamma_g^2}
\int_0^{4K(k^2)/T_2^{\prime}} \mathrm{d}t^{\prime} \, \mathbfit{B}^{\prime} \bmath{\cdot} 
\frac{\mathrm{d}\mathbfit{M}^{\prime}}{\mathrm{d}t^{\prime}}
\ee
where $V$ is the grain volume. To first order in $T_2^{\prime}$,
\begin{multline}
\mathbfit{B}^{\prime} \bmath{\cdot} \frac{d\mathbfit{M}^{\prime}}{dt^{\prime}} =
\left( c_2^2 - k^2 c_1^2 -c_3^2 \right) \mathrm{sn}(T^{\prime}_2 t^{\prime}, k^2)
\mathrm{cn}(T^{\prime}_2 t^{\prime}, k^2) \mathrm{dn}(T^{\prime}_2 t^{\prime}, k^2)
+ T_2^{\prime} \left[ k^2 \left( c_2^2 - c_3^2 \right)
\mathrm{sn}^2(T^{\prime}_2 t^{\prime}, k^2) \mathrm{cn}^2(T^{\prime}_2 t^{\prime}, k^2)
\right. \\
\left. + \left( c_2^2 - k^2 c_1^2 \right)
\mathrm{sn}^2(T^{\prime}_2 t^{\prime}, k^2) \mathrm{dn}^2(T^{\prime}_2 t^{\prime}, k^2)
+ \left( k^2 c_1^2 + c_3^2 \right)
\mathrm{cn}^2(T^{\prime}_2 t^{\prime}, k^2) \mathrm{dn}^2(T^{\prime}_2 t^{\prime}, k^2)
\right] .
\end{multline}
The zeroth-order term integrates to zero identically. 
The first-order term then yields 
\be
\int_0^{4 K(k^2)/T_2^{\prime}} dt^{\prime} \, \mathbfit{B}^{\prime} \bmath{\cdot} \frac{d
\mathbfit{M}^{\prime}}{dt^{\prime}} = \frac{4 \left\{ z_1 [E(k^2) + (k^2-1)
K(k^2)] + k^2 z_2 E(k^2) \right\} (q-1)}{3 k^2 (r_2-1)^2 (r_3-1) (r_3-q)}
+ O(T_2^{\prime})
\ee
where $E(k^2)$ is the complete elliptic integral of the second kind,
\be
E(k^2) = \int_0^{\pi/2} \mathrm{d}\theta (1-k^2 \sin^2 \theta)^{1/2} ,
\ee
\be
z_1 = 2 (r_3-r_2) - r_3^2 (r_2-1) + r_2^2 (r_3-1) ,
\ee 
and
\be
z_2 = - (r_3-r_2) +2 r_3^2 (r_2-1) + r_2^2 (r_3-1) .
\ee
Thus,
\be
A(q) = \frac{\mathrm{d}q}{\mathrm{d}t} = - \tau_{\mathrm{int}}^{-1} \, 
\frac{\left\{ z_1 [P(k^2) + k^2 -1] + k^2 z_2 P(k^2) \right\} (q-1) (r_3-q)}
{3k^2 (r_3-1)} 
\label{eq:drift-low-q}
\ee
where $P(k^2) = E(k^2)/K(k^2)$ and
\be
\tau_{\mathrm{int}} = \frac{\gamma_g^2 I_1^3}{2 \chi_0 V T_2 J^2} .
\ee

In the limit $r_3 \rightarrow r_2$, $k^2 \rightarrow 0$, $z_1 \rightarrow 0$,
$z_2 \rightarrow 3 r_2^2 (r_2-1)$, and
\be
\frac{z_1 [P(k^2) + k^2 -1] + k^2 z_2 P(k^2)}{k^2} \rightarrow z_2 .
\ee
Thus,
\be
A(q) \rightarrow - \frac{2 \chi_0 V T_2 J^2}{\gamma_g^2 I_1 I_2^2} \
(q-1) (r_2 -q)
\ee
which reproduces  equation (\ref{eq:drift-symm}) for a symmetric grain when 
$\omega_{\mathrm{rot}} T_2 \ll 1$. 

In the limit $q \rightarrow 1$, 
\be
A(q) \rightarrow - \tau_{\mathrm{int}}^{-1} \, 
\frac{r_3^2 (r_2-1) + r_2^2 (r_3-1)}{2} \ (q-1) .
\ee
Note that this expression also agrees with the result for a symmetric grain
when $r_2=r_3$.  

As expected, $A(q) \rightarrow 0$ when $q \rightarrow r_2$, though this is
not apparent from inspection of equation (\ref{eq:drift-low-q}).  In this limit,
\be
A(q) \rightarrow - \tau_{\mathrm{int}}^{-1} \,
\frac{2 (r_2-1) (r_3-r_2)}{3 (r_3-1)} \left[ r_3 - r_2 + r_3^2 (r_2-1) + 2 r_2^2 
(r_3-1) \right] \left\{ \ln \left[ \frac{(r_2-1) (r_3-r_2)}{(r_3-1) |r_2-q|} 
\right] \right\}^{-1} .
\label{eq:limit-r2}
\ee
Thus, $A(q)$ falls off much faster with $r_2-q$ than for a symmetric grain
as $q \rightarrow r_2$.  With the absolute value of $r_2-q$ in equation
(\ref{eq:limit-r2}), the expression is correct in the $q \rightarrow r_2$
for both $q<r_2$ and $q>r_2$.  

\subsection{The case that $r_2 < q < r_3$}

When $r_2 < q < r_3$, a derivation analogous to that above yields 
\be
A(q) = - \tau_{\mathrm{int}}^{-1} \, 
\frac{\left\{ z_2 [P(k^{-2}) + k^{-2} -1] + k^{-2} z_1 P(k^{-2}) \right\} (q-1) 
(r_3-q)}{3k^{-2} (r_3-1)} .
\label{eq:drift-high-q}
\ee
In the limit $q \rightarrow r_3$, 
\be
A(q) \rightarrow - \tau_{\mathrm{int}}^{-1} \, 
\frac{r_3 - r_2 + r_2^2 (r_3-1)}{2} \ (r_3-q) .
\ee

\section{The Diffusion Coefficient for Non-Symmetric Grains}
\label{sec:diffusion-coeff}

As for symmetric grains, we use equation (\ref{eq:diff-coeff-as-integral})
to find the diffusion coefficient. The drift coefficient $A(q)$ is given by
equations (\ref{eq:drift-low-q}) and (\ref{eq:drift-high-q}) for the cases that
$q < r_2$ and $q > r_2$, respectively. For later
convenience, define $A_1(q) = - \tau_{\mathrm{int}} A(q)$. 

The thermal-equilibrium distribution function is
\be
f_{\mathrm{TE}}(q) \propto \exp(-b q) \, s^{\prime}(q)
\ee
where
\be
s(q) = 1 - \frac{2}{\pi} \int_0^{\alpha_{\mathrm{max}}} \mathrm{d}\alpha \left[
\frac{r_2 - q + (r_3 - r_2) \cos^2 \alpha}{r_2 -1 + (r_3 - r_2) \cos^2 \alpha}
\right]^{1/2} ,
\ee
\be
\alpha_{\mathrm{max}} =
\begin{cases}
\pi/2 &, q \le r_2 \\
\cos^{-1} \left[ \frac{q - r_2}{r_3 - r_2} \right]^{1/2} &, q > r_2
\end{cases} ,
\ee
and $s^{\prime}(q) = \mathrm{d}s/\mathrm{d}q$ \citep{W09}. 
Thus,
\begin{align}
s^{\prime}(q) &= \pi^{-1} \int_0^{\alpha_{\mathrm{max}}} \mathrm{d}\alpha \left\{ \left[
r_2 -1 + (r_3 - r_2) \cos^2 \alpha \right] \left[ r_2 - q + (r_3 - r_2)
\cos^2 \alpha \right] \right\}^{-1/2} \\
&= \pi^{-1} \int_0^{\alpha_{\mathrm{max}}} \mathrm{d}\alpha \left\{ \left[ r_3 - 1 -
(r_3 - r_2) \sin^2 \alpha \right] \left[ r_3 - q - (r_3 - r_2) \sin^2 \alpha
\right] \right\}^{-1/2}
\label{eq:s-prime-with-sines}
\end{align}
except when $q = r_3$, for which
\be
s^{\prime}(r_3) = \frac{1}{2} [(r_3 - r_2) (r_3 - 1)]^{-1/2} .
\ee
When $q=1$, 
\be
s^{\prime}(1) = \frac{1}{2} [(r_2 -1) (r_3 -1)]^{-1/2} .
\ee 
After slight manipulation, the expression in equation
(\ref{eq:s-prime-with-sines}) is of the form of integral 2.616\#1 in
\citet{Gradshteyn-Ryzhik15}, yielding
\be
\label{eq:s_prime_low}
s^{\prime}(q) = \pi^{-1} \left[ (r_2 - 1) (r_3 - q) \right]^{-1/2} K(k^2)
\ee
when $q < r_2$.
Since $K(k^2) \rightarrow \pi/2$ as $k \rightarrow 0$, 
$s^{\prime}(q) \rightarrow \frac{1}{2} [ (r_2 -1) (r_2 - q)]^{-1/2}$ as 
$r_3 \rightarrow r_2$, reproducing the result for a symmetric grain
\citep[see eq.~13 in][]{W09}.
A change of variables $x = (r_3 - r_2) \sin^2 \alpha$ puts the expression in
equation (\ref{eq:s-prime-with-sines}) into the form of integral 3.147\#3 in
\citet{Gradshteyn-Ryzhik15}, yielding
\be
s^{\prime}(q) = \pi^{-1} \left[ (r_3 - r_2) (q-1) \right]^{-1/2} K \left( k^{-2}
\right)
\label{eq:s_prime_high}
\ee
when $q > r_2$. As for a symmetric grain, $s^{\prime}(q) \rightarrow \infty$
as $q \rightarrow r_2$. 

Defining $G(q) = A_1(q) s^{\prime}(q)$, equation (\ref{eq:diff-coeff-as-integral})
becomes
\be
\label{eq:D1}
\tau_{\mathrm{int}} D(q, b) = \frac{\exp(bq)}{s^{\prime}(q)} \left[ 
\tau_{\mathrm{int}} D(1, b) \exp(-b) s^{\prime}(1) - 2 \int_1^q G(q^{\prime})
\exp(-bq^{\prime}) \mathrm{d}q^{\prime} \right] .
\ee

From equations (\ref{eq:drift-low-q}) and (\ref{eq:s_prime_low}), 
\be
\label{eq:g_low}
G(q) = \frac{(r_2 - 1)^{1/2} (r_3 - q)^{3/2} \left[ \left( z_1 + k^2 z_2 \right)
E(k^2) - z_1 (1 - k^2) K(k^2) \right]}{3 \pi (r_3 -1) (r_3 - r_2)}
\ee
when $q < r_2$. Similarly, from equations (\ref{eq:drift-high-q}) and
(\ref{eq:s_prime_high}),
\be
\label{eq:g_high}
G(q) = \frac{(r_3 - r_2)^{1/2} (q-1)^{3/2} \left[ \left( z_2 + k^{-2} z_1 \right)
E(k^{-2}) - z_2 (1-k^{-2}) K(k^{-2}) \right]}{3 \pi (r_3 -1) (r_2 - 1)}
\ee
when $q > r_2$.

\subsection{The boundary at $q=r_2$}

With the above results, we are now prepared to demonstrate that the boundary
at $q=r_2$ is not a natural boundary. Recall that $A(q)$, $D(q)$, and
$\mathrm{d}D/\mathrm{d}q$ must all vanish at $q=r_2$ in order for it to be a
natural boundary.
Of course, $A(q)$ vanishes at $q=r_2$. As noted above,
$s^{\prime}(q) \rightarrow \infty$ as $q \rightarrow r_2$. Since
the integral in equation (\ref{eq:D1}) does not diverge, $D(q)$ also
vanishes at $q=r_2$. To see that the integral does not diverge, it is
sufficient to note that the function $G(q)$ does not diverge anywhere.
This function vanishes
as $q \rightarrow 1$ and $q \rightarrow r_2$ and
\be
G(q) \rightarrow \frac{\left[ (r_2 - 1) (r_3 - r_2) \right]^{1/2} (z_1 + z_2)}
{3 \pi (r_3 - 1)}
\ee
as $q \rightarrow r_2$ (from both below and above). As $q \rightarrow r_2$,
$k^2 \rightarrow 1$, $E(k^2) \rightarrow 1$, and
$K(k^2) \rightarrow -\frac{1}{2} \ln (1-k^2)$. Thus,
the term $(1-k^2) K(k^2)$ in $G(q)$ does not diverge; rather, it vanishes as
$q \rightarrow r_2$. 

The derivative
\be
\frac{\mathrm{d}(\tau_{\mathrm{int}} D)}{\mathrm{d}q} (q=r_2) = - \frac{2 G(r_2)}
{s^{\prime}(r_2)} + \frac{b \exp(br_2) H(r_2)}{s^{\prime}(r_2)} - \frac{\exp(br_2)
H(r_2) s^{\prime \prime}(r_2)}{[s^{\prime}(r_2)]^2} 
\label{eq:deriv1}
\ee
where $H(q)$ is the quantity in square brackets in equation (\ref{eq:D1}), 
\be
\label{eq:h-of-q}
H(q) = \tau_{\mathrm{int}} D(1, b) \exp(-b) s^{\prime}(1) - 2 \int_1^q G(q^{\prime})
\exp(-bq^{\prime}) dq^{\prime} . 
\ee
Since $s^{\prime}(q)$ diverges at $q=r_2$ and $G(q)$ does not, the first term on
the right-hand side in equation (\ref{eq:deriv1}) vanishes. Since $D(q)$
must be non-negative for all $q$, so must $H(q)$. Since $G(q)$ is
non-negative, $H(q)$ can, at most, vanish only at $q=r_3$. Thus, the numerator
in the second term on the right-hand side in equation (\ref{eq:deriv1}) is
non-zero but does not diverge, implying that the second term vanishes.
From equations (\ref{eq:s_prime_low}) and
(\ref{eq:s_prime_high}) and the relation 
\be
\frac{\mathrm{d}K(k^2)}{\mathrm{d}k^2} = \frac{E(k^2) - (1-k^2) K(k^2)}{2 k^2
(1-k^2)} ,
\ee
\be
s^{\prime \prime}(q) \rightarrow \pm \frac{r_3 -1}{2 \pi [ (r_2-1) (r_3 -r_2)]^{3/2}}
\ \frac{1}{1-k^{\pm 2}}
\ee
as $q \rightarrow r_2$; the + (-) sign applies when $q<r_2$ ($q > r_2$).
Thus, the final term on the right-hand
side in equation (\ref{eq:deriv1}) approaches $\mp \infty$ as
$q \rightarrow r_2$ from below (minus sign) or above. That is,
$\mathrm{d}D/\mathrm{d}q$
diverges at $q=r_2$ and this point is not a natural boundary. 

\subsection{The constant of integration}

As for symmetric grains, we demand that $D(q,b) \rightarrow 0$ as
$b \rightarrow 0$. Thus, it must be that $H(q=r_3, b)$ falls off at least
as rapidly as $\exp(-br_3)$. Of course, $H(q=r_3, b) \equiv 0$
satisfies this condition. This choice corresponds to that made by
\citet{W09} for symmetric grains, i.e.~$C(b) \equiv 0$. As
$q \rightarrow r_3$,
\be
s^{\prime \prime}(q) \rightarrow - \frac{1}{4} \left[ (r_3 - r_2) (r_3 -1)
\right]^{-1/2} (r_2 - 1)^{-1} .
\ee
Since $s^{\prime}(r_3)$ and $s^{\prime \prime}(r_3)$ are both non-zero and
non-divergent and $G(r_3) = 0$, if $H(r_3) = 0$ then $q=r_3$ is a natural
boundary.

\section{Simulations of Barnett relaxation for non-symmetric grains}
\label{sec:simulations}

Lacking a first-principles model of Barnett relaxation, we will adopt
$H(r_3) \equiv 0$ for simplicity. Equation (\ref{eq:D1}) becomes
\be
\tau_{\mathrm{int}} D(q, b) = \frac{2 \exp(bq)}{s^{\prime}(q)} \int_q^{r_3}
G(q^{\prime}) \exp(-bq^{\prime}) \mathrm{d}q^{\prime} .
\ee

From the definition of the inertia tensor, $r_3$ can be arbitrarily large
but it must be that $r_2 \le r_3/(r_3-1)$. In order to check the thermal
flipping rate, we ran simulations of Barnett relaxation for three grain
shapes. Shapes 1, 2, and 3 have $(r_2, r_3)$ = $(1.3, 1.5)$, $(1.49, 1.5)$, 
and $(1.01, 1.5)$, respectively.

The Langevin equation is
\be
\mathrm{d}q = A(q) \, \mathrm{d}t + \sqrt{D(q)} \, \mathrm{d}w
\ee
where $\mathrm{d}t$ is a time step and $\mathrm{d}w$ is a Gaussian random
variable with variance $\mathrm{d}t$. Adopting dimensionless quantities
$\mathrm{d}t^{\prime} = \mathrm{d}t/\tau_{\mathrm{int}}$,
$\mathrm{d}w^{\prime} = \mathrm{d}w/\sqrt{\tau_{\mathrm{int}}}$, 
$B_1(q) = \sqrt{\tau_{\mathrm{int}} D(q)}$, and recalling the definition
$A_1(q) = - \tau_{\mathrm{int}} A(q)$, the Langevin equation becomes
\be
\mathrm{d}q = - A_1(q) \, \mathrm{d}t^{\prime} + B_1(q) \, \mathrm{d}w^{\prime} .
\ee

\begin{figure}
\begin{minipage}{9.1cm}
\includegraphics[width=90mm]{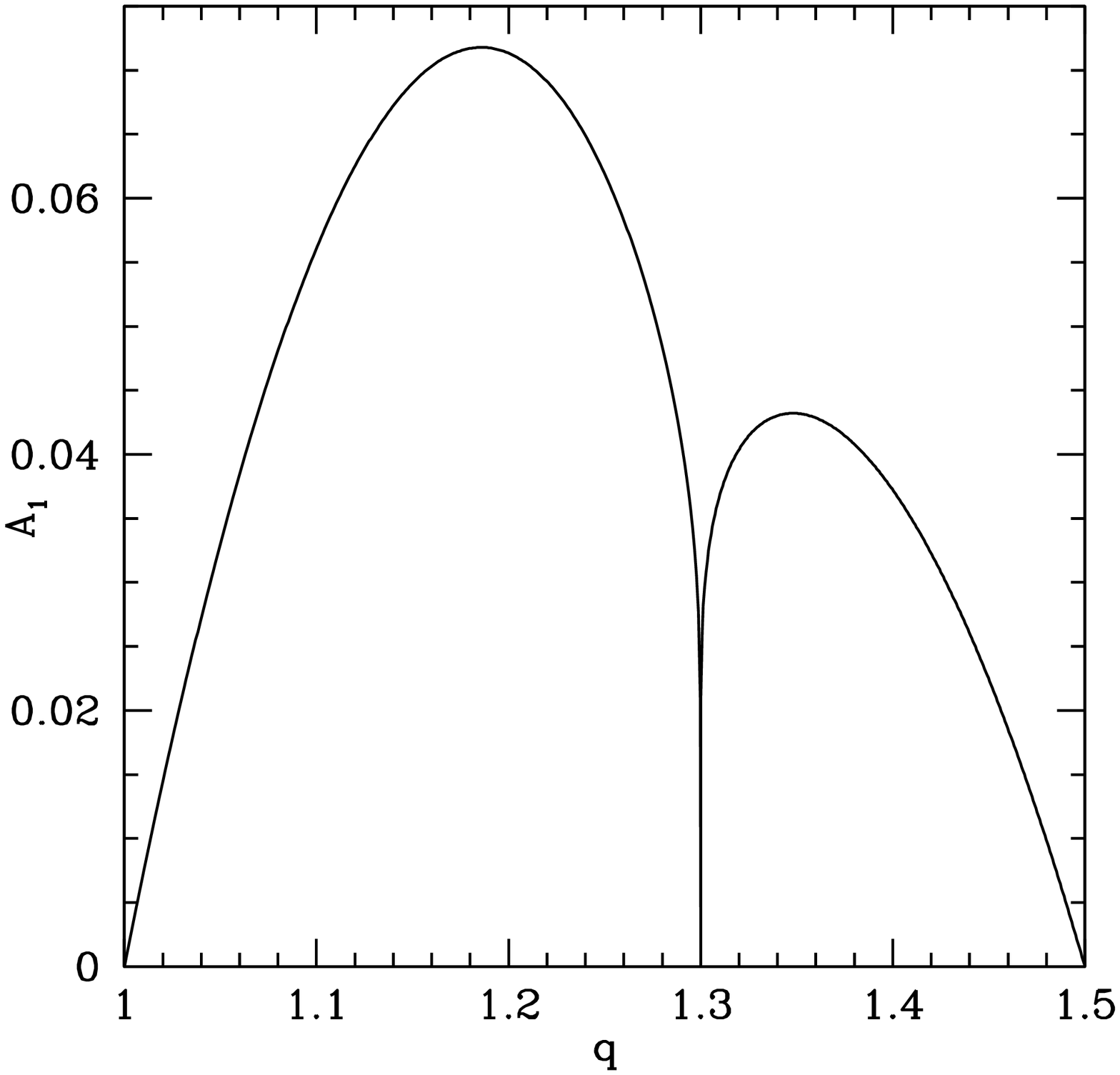}
\end{minipage}
\begin{minipage}{9.1cm}
\includegraphics[width=90mm]{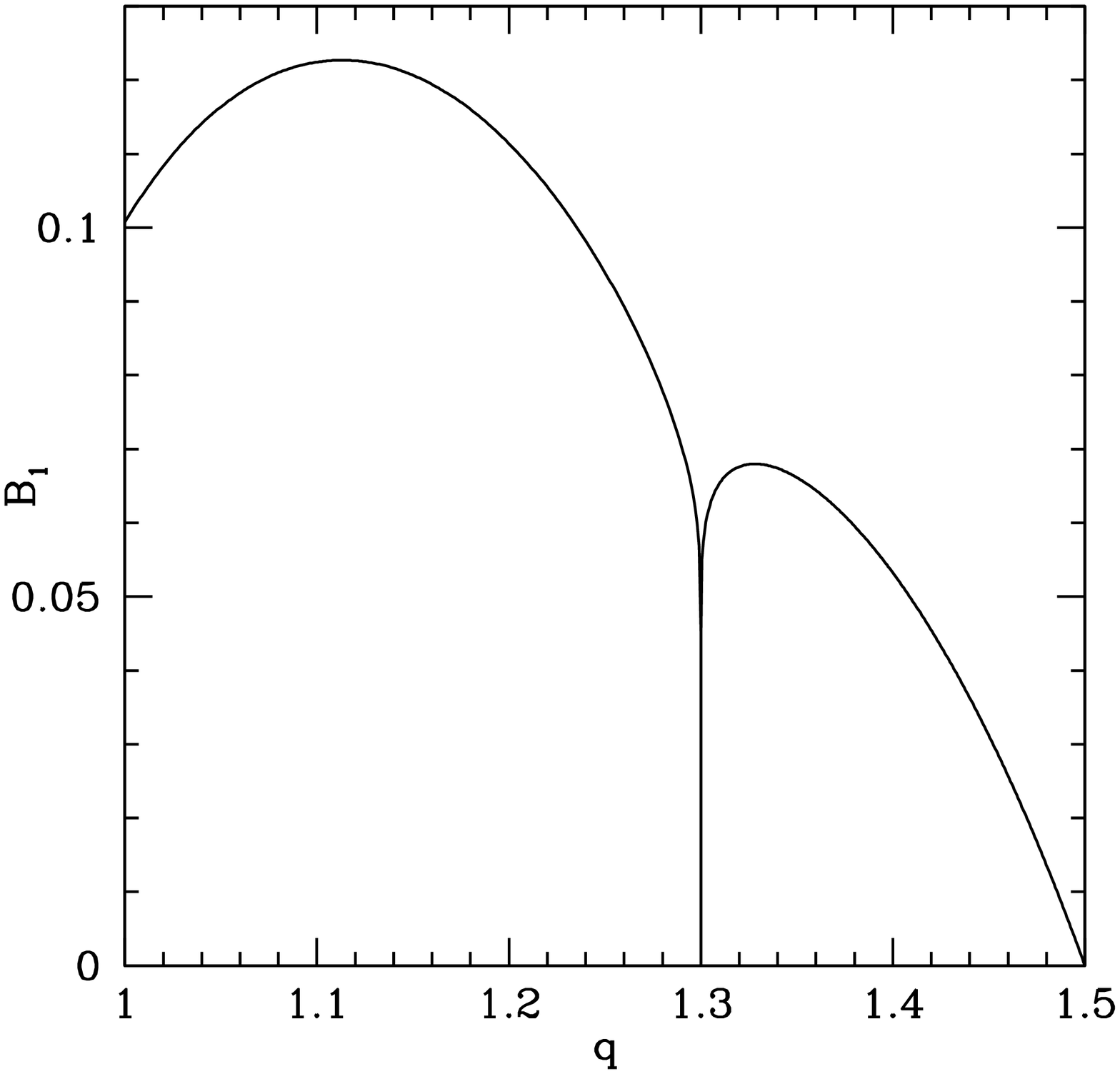}
\end{minipage}
\caption{
{\it Left:}
$A_1(q) = - \tau_{\mathrm{int}} A(q)$, where $A(q)$ is the drift coefficient,
for shape 1 ($r_2 = 1.3$, $r_3 = 1.5$).
{\it Right:}
$B_1(q) = \sqrt{\tau_{\mathrm{int}} D(q)}$, where $D(q)$ is the diffusion 
coefficient, for shape 1 and $b = 10.0$.
        }
\label{fig:a1-b1}
\end{figure}

For each grain shape, we consider $b = 1.0$, 5.0, 10.0, and 20.0. For
each simulation, we first construct interpolation tables for $A_1(q)$
and $B_1(q)$. As seen in Fig. \ref{fig:a1-b1}, these 
functions drop steeply to zero as $q \rightarrow r_2$, where flipping occurs.
Thus, we tabulate in parameter $\log (r_2 -q)$ or $\log(q - r_2)$ when
$q < r_2$ or $q > r_2$, respectively, rather than in $q$. The tables 
contain values for $2 \times 10^4$ values of $q$ (half with $q < r_2$ and
half with $q > r_2$) computed using {\sc Mathematica}. 
We also ran a simulation with $2 \times 10^3$
values of $q$ in the interpolation tables and found that the results were
not significantly affected. 

The simulations adopt a step size $\mathrm{d}t^{\prime} = 10^{-5}$. We also ran a
simulation with step size $\mathrm{d}t^{\prime} = 10^{-6}$ and found that the
average flip time was not substantially affected. If a step causes the grain to 
overstep a boundary (i.e.~$q < 1$ or $q > r_3$), then $\mathrm{d}t^{\prime}$ is 
reduced by a factor of 10 and the step is attempted again. Random numbers and
the Gaussian random variables $\mathrm{d}w$ are computed using modified
versions of the routines {\sc ran2} and {\sc gasdev} from \citet{Press92}.
As seen in Fig. 4 of \citet{WD03}, a grain with $q$ slightly larger than
$r_2$ can naturally transition into either the positive or negative flip
state with respect to $\bmath{\hat{a}}_1$ if $q$ changes to a value somewhat
less than $r_2$. Which flip state results depends on the phase of the grain
rotation at the moment when $q$ crosses $r_2$. Half of the total phase
corresponds to a transition to the positive flip state and the other half
to the negative flip state. Thus, whenever $q$ crosses $r_2$
from above, the flip state with respect to $\bmath{\hat{a}}_1$ is chosen
randomly, with equal probability for the positive and negative flip states.
(We do not bother to track the flip states with respect to
$\bmath{\hat{a}}_3$ when $q > r_2$.)

\begin{figure}
\begin{minipage}{9.1cm}
\includegraphics[width=90mm]{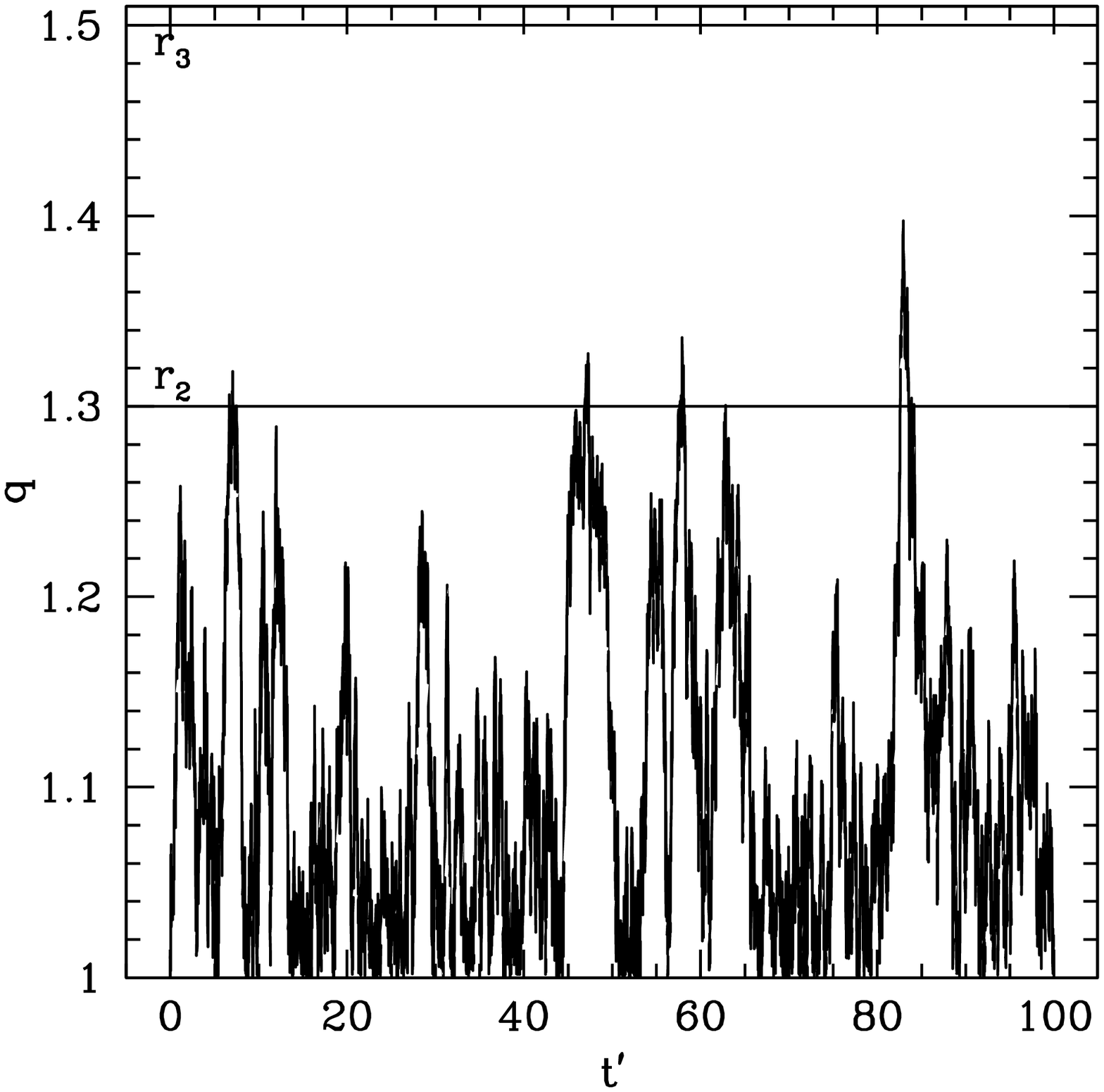}
\end{minipage}
\begin{minipage}{9.1cm}
\includegraphics[width=90mm]{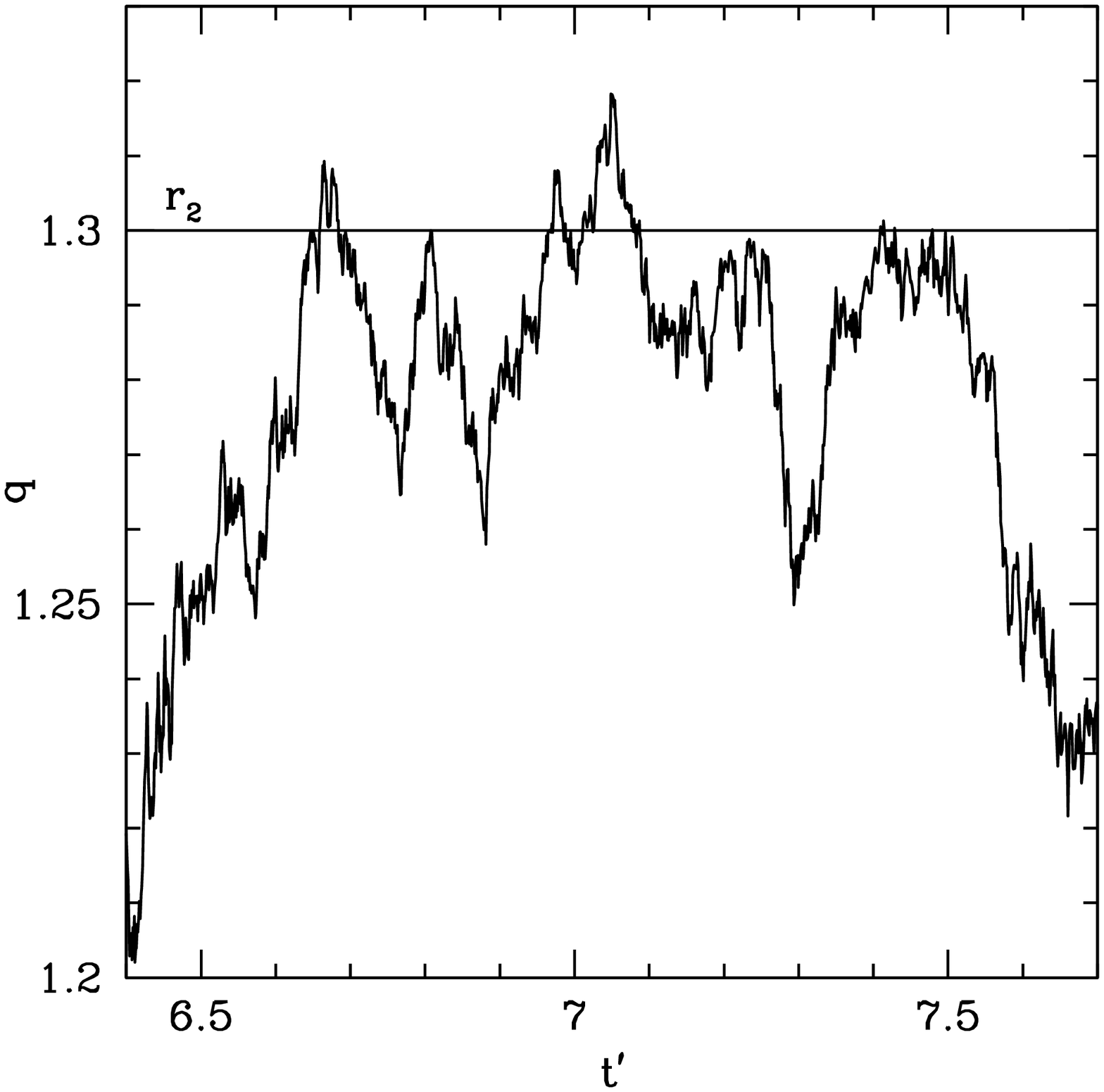}
\end{minipage}
\caption{
{\it Left:}
Simulation of Barnett relaxation ($q$ versus $t^{\prime} = t/\tau_{\mathrm{int}}$)
for shape 1 ($r_2 = 1.3$, $r_3 = 1.5$)  with $b=10$.
{\it Right:}
A zoom-in on the region around a potential flip.
        }
\label{fig:flips_b10}
\end{figure}

Fig. \ref{fig:flips_b10} shows the results of a simulation for shape 1
with $b = 10.0$. A flip occurs when the grain starts in one flip state with
respect to $\bmath{\hat{a}}_1$ with $q < r_2$, evolves to $q > r_2$, and
evolves back to $q < r_2$ in the opposite flip state with respect to
$\bmath{\hat{a}}_1$. The right panel of 
Fig. \ref{fig:flips_b10} zooms in on the first opportunity for a flip
to occur, when $t^{\prime} \approx 7$. Not surprisingly, there are multiple
crossings of $q=r_2$ in rapid succession, with $q$ remaining near $r_2$.
It would not be sensible to count each of these crossings as a potential
flip. Instead, we adopt the following 
criterion for a genuine flip to occur: the grain must cross $q=r_2$ from 
below and later (perhaps after multiple crossing of $q=r_2$) return to 
$q = \frac{1}{2} (1 + r_2)$ in the opposite flip state.

We ran long simulations to find the average time between flips, 
$\tau_{\mathrm{flip}}$, for the three shapes and four values of $b$ indicated
above. The results for 
$\tau_{\mathrm{flip}}^{\prime} = \tau_{\mathrm{flip}} / \tau_{\mathrm{int}}$ 
are summarized in Table \ref{tab:t-flip-int}.
Each shape exhibits rapid thermal flipping for sufficiently low $b$. As
the grain shape approaches that of an oblate symmetric grain ($r_2 = r_3$),
the flipping rate decreases. This is as expected, since we chose the
constant of integration for the diffusion coefficient to be consistent with
the choice in \citet{W09}. For shape 3, with $r_2$ close to 1 and
$r_3$ considerably larger, the flipping rate is relatively fast. In this
case, so long as $b$ is not too large, the grain spends much or most of the
time with $q > r_2$, i.e.~in a flip state with respect to
$\bmath{\hat{a}}_3$.

For comparison, the last column in Table \ref{tab:t-flip-int} is 
$\tau_{\mathrm{flip}}^{\prime}(\mathrm{LD99}) = \{ \exp [- (b-1)/2] \}^{-1}$,
the approximate expression for the flip time from \citet{LD99a}.

\begin{table}
\begin{minipage}{16cm}
\caption{Average time between flips in stochastic simulations
\label{tab:t-flip-int}}
\begin{tabular}{@{}lllll}
\hline
Shape & $b$ & $\tau_{\mathrm{flip}}^{\prime}$ & Number of flips & 
$\tau_{\mathrm{flip}}^{\prime}(\mathrm{LD99})$\\
\hline
1 & 1.0  & 13.96 & 286438 & 1.0 \\
1 & 5.0  & 21.27 & 2960253  & 7.4 \\
1 & 10.0 & 48.39 & 289085  & 90 \\
1 & 20.0 & 448.3 & 187133   & $1.4 \times 10^4$ \\
2 & 1.0  & 109.4 & 54834 & 1.0 \\
2 & 5.0  & 259.4 & 46207 & 7.4 \\
2 & 10.0 & 1272	 & 49398 & 90 \\
2 & 20.0 & $6.942 \times 10^4$ &  1949 & $1.4 \times 10^4$ \\
3 & 1.0  & 0.3350 & 17896669 & 1.0 \\
3 & 5.0  & 0.4254 & 28184360 & 7.4 \\
3 & 10.0 & 0.5869 & 71503365 & 90 \\ 
3 & 20.0 & 0.9727 & 86290047 & $1.4 \times 10^4$ \\
\hline
\end{tabular}
\end{minipage}
\end{table}

\section{Summary}
\label{sec:summary}

The following are our principal results.

1. We revisited Barnett relaxation in grains with dynamic symmetry,
concluding that thermal flipping (i.e.~flipping induced by internal
relaxation) does not occur. We do not agree with \citet{W09} that
$q=r_2$ must be a natural boundary, although it could be depending on the
value of the unknown constant of integration in equation
(\ref{eq:diff-coeff-as-integral}). 
In any case, the probability that a grain reaches exactly $q=r_2$,
required for a flip to occur, is infinitesimal. 

2. We derived expressions for the Barnett dissipation rate for grains lacking
dynamic symmetry (equations \ref{eq:drift-low-q} and \ref{eq:drift-high-q})
in the limit of low rotational frequency (i.e.~the spin-spin relaxation
time is much less than the rotation period), 
assuming that the spin-spin and spin-lattice relaxation times are equal.

3. Given the above dissipation rates, we derived expressions for the diffusion
coefficient (equations \ref{eq:D1}--\ref{eq:g_high}),
which involve a constant of integration whose behavior in the low-temperature
limit ($b \rightarrow \infty$) is constrained. 

4. We show that the boundary at $q=r_2$ is not a natural boundary for grains
lacking dynamic symmetry. Thus, thermal flipping does occur in these grains.

5. We present results of long-time simulations of the internal grain dynamics
using the Langevin equation for a few non-symmetric grain shapes. As expected,
flipping does occur. For these simulations, we chose the constant of
integration for the diffusion coefficient such that $q=r_3$ is a natural
boundary. A first-principles analysis of Barnett relaxation is needed to
definitively set the value of this constant. 
We also provide a practical definition of a flip.

Order-of-magnitude estimates of the spin-spin relaxation times are
$T_2 \sim 3 \times 10^{-11}$ to $3 \times 10^{-9} \, \mathrm{s}$
for Barnett relaxation \citep{Draine96} and
$T_2 \sim 3 \times 10^{-5}$ to $3 \times 10^{-4} \, \mathrm{s}$ for nuclear
relaxation \citep{LD99b}.
Thus, our assumption of the low-frequency limit is much more severe for
nuclear relaxation, limiting the applicability of our quantitative results
to cases with low grain rotational speeds. We adopted the low-frequency
limit in order to exploit analytical solutions to the modified Bloch
equations (\ref{eq:modified-bloch-2})--(\ref{eq:modified-bloch-1}).
In future work, we will numerically integrate these
equations to obtain results valid for all frequencies. We expect that the
dissipation rate will be lower than the results derived here
but that the main conclusion of this paper, namely the possibility of thermal
flipping, will still hold. 

We do not expect our assumption that $T_1 = T_2$ to introduce significant
error in the low-frequency limit, where saturation effects are negligible.
Although $T_1$ and $T_2$ differ by several orders of magnitude for
Barnett relaxation, the low-frequency limit breaks
down only at relatively high rotational frequencies for which thermal
flipping does not occur. In the \citet{LD99b} model of nuclear relaxation,
the time-scale for
spin-spin relaxation within the system of nuclear spins is of roughly the same
magnitude as the time-scale for exchange between the nuclear and electron spin
systems. The time-scale on which the electron spin system couples with the
lattice is much shorter than the time-scale on which the nuclear and electron
spin systems couple with each other, which in turn is much shorter than the
time-scale on which the nuclear spin system couples directly with the lattice.
Thus, the interaction between the nuclear spin system and the lattice is
mediated by the electron spin system and the nuclear spin-lattice time-scale
$T_1$ is of roughly the same magnitude as the nuclear spin-spin time-scale
$T_2$.

In future work, we will extend our stochastic simulations to include the
effects of systematic torques in order to quantify the impact of thermal
trapping. 

\section*{Acknowledgements}

We are grateful to Bruce Draine and Alex Lazarian for helpful discussions.

\bibliographystyle{mnras}
\bibliography{mybib} 

\bsp	
\label{lastpage}
\end{document}